\documentclass[pra,showpacs,twocolumn,superscriptaddress,amsmath,amssymb]{revtex4}

\newcommand{\beq}{\begin{equation}}
\newcommand{\eeq}{\end{equation}}

\usepackage{graphicx}
\usepackage{color}
\usepackage{dcolumn,amsmath}

\begin{document}

\title{External field control of spin-dependent rotational decoherence of ultracold polar molecules}

\author{Alexander Petrov$^{a,b}$, Constantinos Makrides$^{a,c}$, and Svetlana Kotochigova$^{a}$${^\ast}$
               \thanks{$^\ast$Corresponding author. Email: skotoch@temple.edu} \\
               \vspace{6pt}
          $^{a}${\em Department of Physics, Temple University, Philadelphia, PA 19122-6082, USA}\\
          $^{b}${\em St. Petersburg Nuclear Physics Institute, Gatchina, 188300; 
                     Division of Quantum Mechanics, St. Petersburg State University, 198904, Russia}\\
          $^{c}${\em Department of Physics and Astronomy, University of Toledo, Mailstop 111,
                     Toledo, Ohio 43606, USA }
}

\begin{abstract}
We determine trapping conditions for ultracold polar molecules, where
pairs of internal states experience identical trapping potentials. Such
conditions could ensure that detrimental effects of inevitable
inhomogeneities across an ultracold sample are significantly reduced. In
particular, we investigate the internal rovibronic and hyperfine quantum
states of ultracold fermionic ground-state $^{40}$K$^{87}$Rb polar
molecules, when static magnetic, static electric, and trapping laser
fields are simultaneously applied. Understanding the effect of changing
the relative  orientation or polarization of these three  fields is of
crucial importance for creation of decoherence-free subspaces built from
two or more  rovibronic states.  Moreover, we evaluate the
induced dipole moment of the molecule in the presence of these fields,
which will allow control of interactions between molecules in different
sites of an optical lattice and study the influence of the interaction
anisotropy on the ability to entangle polar molecules.
\end{abstract}

\maketitle


\section{Introduction}
The experimental realization of a high phase-space density, quantum-degenerate gas of 
molecules, prepared in a single quantum state, \cite{Ni2008,Ospelkaus2009,Danz2008,Danz2010}
opens up exciting prospects for the ultimate control 
of their internal and external degrees of freedom. 
In addition, significant progress has been made in loading and manipulating diatomic molecular
species in periodic optical potentials \cite{Ospelkaus2010,Miranda2011,Danz_Science2008}.  
Polar molecules are of particular interest in such experiments as they have permanent 
electric dipole moments and therefore can interact via long-range tunable dipole-dipole interactions. 
Trapped in an optical lattice these molecules can form new types
of highly-correlated quantum many-body states \cite{Baranov2008,Lahaye2009}.
Moreover, it has been proposed \cite{DeMille2002} that they can be quantum
bits of a scalable quantum computer.  Finally, ultracold molecules are also promising systems 
to perform high-precision measurements of a possible time variation of fundamental physical
constants.
In parallel, there is growing interest orienting (non-degenerate) polar molecules using intense pulsed AC fields
sometimes combined with external static electric field \cite{Friedrich1991,Rost1992,Friedrich1995}.
The feasibility of orienting rotationally cold polar molecules in an external field has
been demonstrated \cite{Sakai2003,Nielsen2012}. 

Molecules have complex vibrational, rotational and hyperfine internal structure with 
many internal degrees of freedom \cite{Aldegunde2008,Aldegunde2009}. As was shown in recent experiments \cite{Ospelkaus2010,Neyenhuis2012}, 
coherent control over internal quantum states of molecules plays  a key role in manipulation of molecules with
a long coherence time. An important property for controlling a molecule with light fields is its complex molecular 
dynamic polarizability $\alpha(h\nu,\vec{\epsilon})$ at radiation frequency $\nu$
and polarization $\vec{\epsilon}$ ($h$ is Planck's constant). Multiplied
with the laser intensity its real part determines the strength of a
lattice potential. As different internal states have different
polarizability their lattice depths or Stark shifts differ.  

A second important property of polar molecules is its permanent 
dipole moment.  Their rotational levels can be shifted and mixed with one another by applying an external electric
field.  In the presence of both a static external electric field and laser 
fields the ground state has an anisotropic polarizability~\cite{Ospelkaus2009}.
The anisotropy of the dynamic polarizability of these levels manifests itself as a dependence  on the relative
orientation of the polarization of the trapping laser and the DC electric
field.  Finally, alkali-metal polar molecules have a nonzero nuclear electric-quadrupole and  nuclear-magnetic moments
of the constituent atoms.
Then by applying a magnetic field, quadrupole and Zeeman interactions further mix states.
The combined action of these three fields can be a powerful tool with which
to manipulate and control ultracold molecules trapped in an optical
potential.

For many applications of ultracold polar molecules it is advantageous
or even required that two or more molecular rotational-hyperfine
states have the same spatial trapping potentials.  This is
as a so-called ``magic''  condition. In atomic gasses magic
conditions occur for specific off-resonant laser frequencies
\cite{Ye2008,Lundblad2010,Derevianko2011}. In molecular systems such
frequencies exist when light resonant or nearly-resonant with molecular
transitions is used \cite{Zelevinsky2008}.  Unwanted spontaneous emission
can then lead to dephasing.  Recent experimental and theoretical studies
\cite{Kotochigova2010,Neyenhuis2012} of ground-state polar molecules
demonstrated that ``magic'' conditions can exist for off-resonant laser
frequencies as long as the angle between the laser polarization and
either an external magnetic or electric field is carefully controlled.
In fact, Ref.~\cite{Neyenhuis2012} showed in measurements of the
AC polarizability and the coherence time for microwave transitions
between rotational states that there exist  a ``magic'' angle between
the orientation of the polarization of the trapping light and a magnetic
field.  In this experiment no electric field was applied.

\begin{figure}
\begin{center}
\includegraphics[width=0.6\textwidth,trim=0 50 0 0,clip]{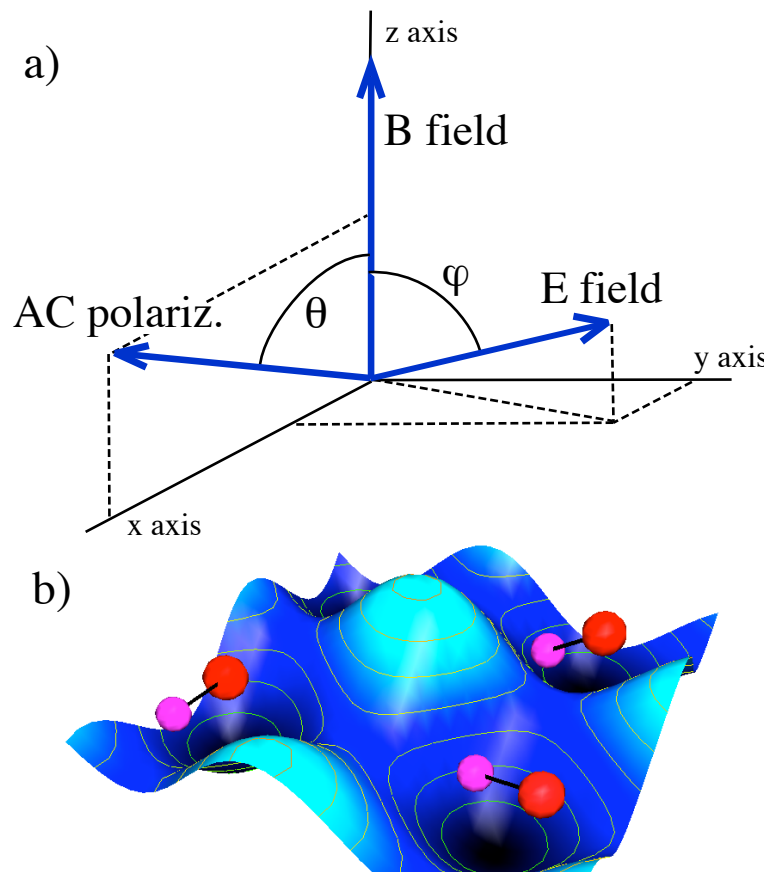}
\end{center}
\caption{(Color online) 
Panel a) shows the orientations of the static electric and magnetic fields  as well as the AC polarization of the optical trapping laser. Without loss of generality, we assume that the magnetic field, $\vec B$, is directed along the $z$ axis, while the polarization $\vec \epsilon$ of the laser lies in the $x$-$z$ plane. The electric field $\vec E$ can be in any direction.
Only two of the three angles that uniquely specify the relative orientations are indicated. Panel b) shows a cartoon of polar molecules held in an optical lattice potential with polarization $\vec\epsilon$.}
\label{scheme}
\end{figure}

Here we extend the ideas of  Ref.~\cite{Aldegunde2008,Aldegunde2009,Kotochigova2010,Neyenhuis2012}
and perform a theoretical study of  the internal rovibronic and
hyperfine quantum states of the KRb  molecules when simultaneously
static magnetic and electric fields as well as nonresonant trapping
lasers are applied.  A schematics is shown in Fig.~\ref{scheme}.
The purpose of the study is to develop a quantitative model for
energy levels, polarizibility, and dipole moments  for an efficient
quantum coherent control of coupled rotational states.  Our research
is closely linked to ongoing experiments with ultracold KRb molecules
\cite{Ospelkaus2010,Miranda2011,Neyenhuis2012}. Understanding the
effect of changing the relative  orientation or polarization of these
three  fields is of crucial importance for creation of decoherence-free
subspaces built from two or more  rovibronic and hyperfine states.

We also evaluate the imaginary part of the polarizability, due to
spontaneous emission from excited electronic states.  Here, the
imaginary part  is calculated assuming that excited vibrational
levels have a linewidth evaluated by either using the
linewidth of atomic  K or Rb or using an
optical-potential approach \cite{Zygelman}.

This paper is set up as follows. In Sec.~\ref{sec:ham} we construct the Hamiltonian
of the rotating KRb molecule in the presence of the three external fields.
We then present results of our calculation of the AC polarizability in Sec.~\ref{sec:reA}.
The dipole-moment is calculated as a function of electric field in Sec.~\ref{sec:dip}.
We finish with a discussion of the imaginary part of the polarizability due to spontaneous emission of electronic excited states
in Sec.~ \ref{sec:ImA}.

\section{Molecular Hamiltonian}\label{sec:ham}

We focus on rotational states of the lowest vibrational level of the
singlet X$^1\Sigma^+$ ground-state potential of $^{40}$K$^{87}$Rb in the presence
of a magnetic and electric field as well as an AC trapping laser field.
We denote the rotational states by the angular momentum quantum number $N$
and  its projection $m_N$ onto the external magnetic field direction.
Both K and Rb have nonzero nuclear spin, which align along the magnetic
field, through the Zeeman interaction.  Nuclear quadrupole interactions
mix these nuclear hyperfine states with the rotation of the molecule.
The relative directions of the fields are defined in Fig.~\ref{scheme}.
Thoughout, angular momentum and tensor algebra is based on Ref.~\cite{Brink}.

In practice we have determined the molecular polarizability and dipole moment 
starting from the molecular basis functions or channels
\begin{eqnarray}
\lefteqn{|N,m_N, m_{\rm a}, m_{\rm b}\rangle\equiv} \\
 && \quad\quad
  \phi_{v=0}(r)\, | {\rm X}^1\Sigma^+ \rangle \,
     Y_{Nm_N}(\alpha\beta)   
  \,| i_{\rm a}m_{\rm a},i_{\rm b}m_{\rm b}\rangle, \nonumber
\end{eqnarray}
where $\phi_{v=0}(r)$ is the $v=0$ radial vibrational wavefunction for
interatomic separation $r$, which for the small $N$ studied here is
to good approximation independent of $N$, and  $|{\rm X}^{1}\Sigma^+
\rangle$ is the electron wavefunction with projections defined along
the internuclear axis.  The spherical harmonic $Y_{Nm_N}(\alpha\beta)$
describes the rotational wavefunction of our $\Sigma$ molecule. The
angles $\alpha$ and $\beta$ and projection $m_N$ are defined with
respect to the magnetic field direction. The nuclear spins ${\vec
\imath}_{\rm a}$ and ${\vec \imath}_{\rm b}$ for atom a and b have projections $m_{\rm a}$
and $m_{\rm b}$ onto the magnetic field.  For $^{40}$K$^{87}$Rb there are $144$
channels  $|N,m_N, m_{\rm a}, m_{\rm b}\rangle$ with $N=0$ and 1.
The degeneracy of states with projections $m_N$ for the same $N$ is
lifted by the interaction between the nuclear quadrupole moment and
the rotation of the molecule~\cite{Ospelkaus2010,Neyenhuis2012}. Here
we focus on hyperfine states whose dominant nuclear spin character is
$m_{\mathrm{a}}=-4,m_{\mathrm{b}}=1/2$ for $^{40}$K and $^{87}$Rb,
respectively.  These states were selected for the experimental
measurements dynamic polarizability in the ground state KRb molecule
\cite{Neyenhuis2012}.

The  effective Hamiltonian for the $v=0$ rotational-hyperfine levels
is given by 
\begin{equation}
H = H_{rot} + H_Z + H_{E} + H_Q + H_{pol} \, , \label{eq:ham}
\end{equation}
and
\begin{eqnarray}
H_{rot} &=& B_{v}\vec {N}^2 \\
H_Z &=& -\sum_{k=a,b} \mu_k \,{\vec \imath}_{k}\cdot\vec B \\
H_E &=& -{\vec d}\cdot \vec{E} \\
H_Q &=& \sum_{k=a,b} Q_k C_2(\alpha\beta)\cdot T_2({\vec \imath}_k,{\vec \imath}_k) \\
H_{pol} &=& -\left(\alpha_{\parallel} {\cal O}_{\parallel} + \alpha_{\perp} {\cal O}_{\perp}\right) I
     \,,
\end{eqnarray}
where $H_{rot}$ is the rotational Hamiltonian with vibrationally-averaged rotational
constant $B_v=\int_0^\infty dr \phi_v(r)\hbar^2/(2\mu r^2) \phi_v(r)$, where $\mu$ is the reduced mass and $\hbar=h/2\pi$. The matrix operator of $H_{rot}$ is diagonal 
with our basis functions. For the KRb dimer the energy spacing $\Delta$ between vibrational
levels $v=0$ and $v=1$ is on the order of $\Delta/h=1500$ GHz, while
$B_v/h=1.1139$ GHz for $v=0$ \cite{Ni2008}.

The next term in the Hamiltonian is the nuclear Zeeman interaction $H_Z$
for each atom, where $\mu_k$ is the nuclear magneton of atom $k$ \cite{Arimondo} and
$\vec B$ is the magnetic field and only affects the nuclear spins.
This contribution is followed by the electric-dipole interaction
$H_E$, which describes the effect of a static electric field
$\vec E$ and contains the vibrationally-averaged molecular dipole moment operator ${\vec d}$.
Matrix elements of $H_E$ follow the realization that $H_E$ can
equivalently be written as $-d_0 \sum_{q=-1}^1 (-1)^q C_{1q}(\alpha\beta)
E_q$, where $d_0=\int_0^\infty dr\, \phi_{v=0}(r) {\cal D}(r) \phi_{v=0}(r)$
and ${\cal D}(r)$ is the $r$-dependent permanent electric dipole moment of the
X$^1\Sigma^+$ potential.
Moreover, $E_q$ are the rank-1 spherical components of $\vec E$ and
$C_{lm}(\alpha\beta)=\sqrt{4\pi/(2l+1)}Y_{lm}(\alpha\beta) $ is a
spherical harmonic of rank $l$.  It follows that matrix elements are
nonzero when $N+1+N'$ is even. We use $d_{v=0}=0.223$ $ea_0$ for KRb,
where $e$ is the electron charge and $a_0=0.05292$ nm is the Bohr radius.
The value is consistent with the result of Ref.~\cite{Ni2008}.

Equation \ref{eq:ham} also includes the nuclear quadrupole interaction
$H_Q$ for each atom. It has coupling constants $Q_k$ and couples the
nuclear spin to rotational states. Here, $T_{2m}({\vec \imath}_k,{\vec
\imath}_k)$ is a rank-2 tensor constructed from  spin ${\vec \imath}_k$.
For KRb the two quadrupole parameters $Q_k$ were first determined in
Ref.~\cite{Ospelkaus2010} based on measurements of transition energies
between sub-levels of the $N=0$ and $N=1$ states.  We use the
more recent values $Q_K/h = 0.452$ MHz and $Q_{Rb}/h = -1.308$ MHz from
\cite{Neyenhuis2012}.

\begin{figure}
\begin{center}
\includegraphics[scale=0.77,trim=9 10 0 0,clip]{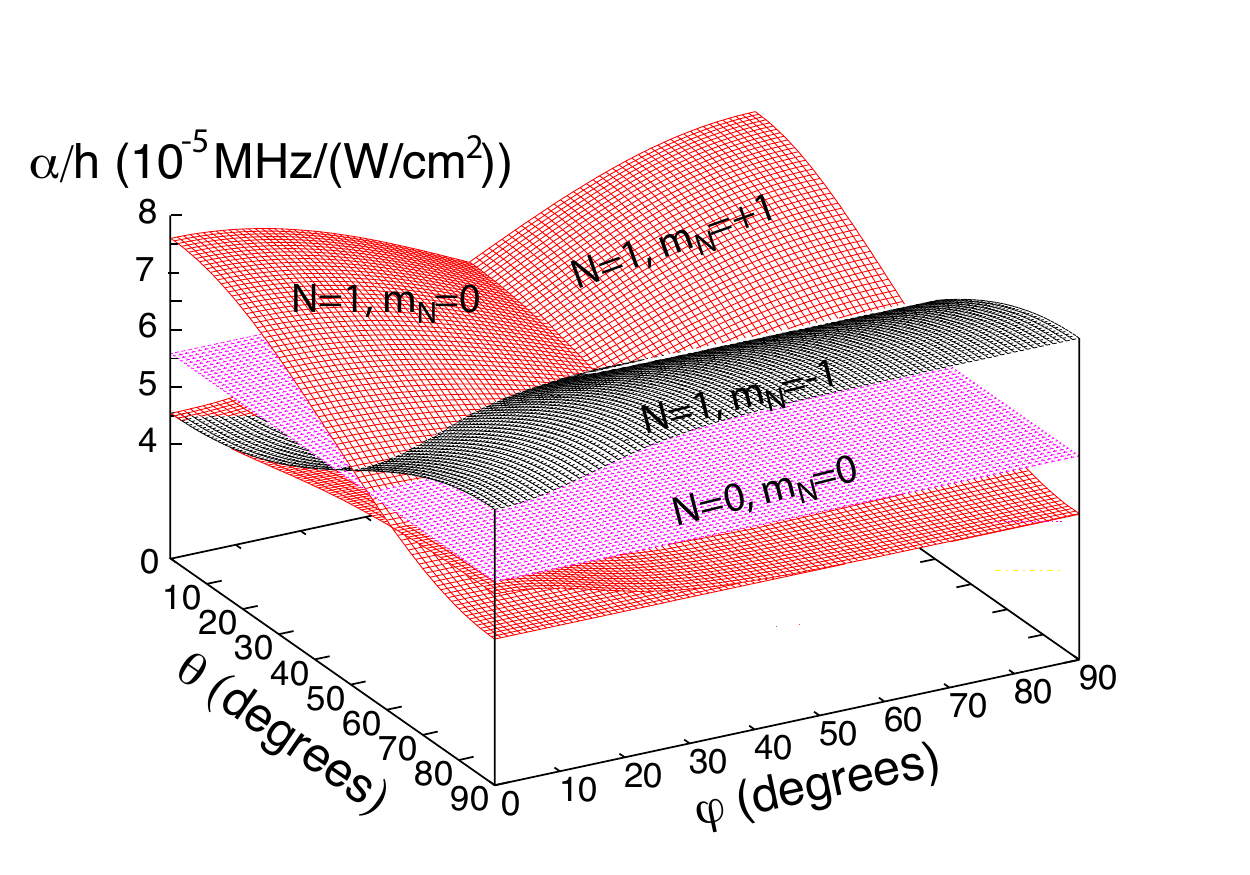}
\end{center}
\caption{(Color online) Polarizability of four rotational-hyperfine states of the $v=0$ vibrational 
level of the X$^1\Sigma^+$ potential of KRb as a function of angle $\theta$ between the polarization 
of the dipole-trap laser and magnetic field, and as a function 
of angle $\varphi$ between the static electric and magnetic fields. 
The electric field lies in the $y$-$z$ plane as defined in Fig.~\ref{scheme}.
The magnetic field strength is $B=545.9$ G, the
electric field strength $E=1$ kV/cm, and the laser at 1063 nm has an intensity of 
$I=2.35$ W/cm$^2$. The four surfaces are labeled by
rotational levels $|N,m_N\rangle=|0,0\rangle$, $|1,0\rangle$, $|1,-1\rangle$, and $|1,+1\rangle$,
respectively.  Their nuclear spin wavefunction is $m_{\rm a}=-4$ and $m_{\rm b}=1/2$ for potassium and rubidium, respectively.
}
\label{polar_all}
\end{figure}

Finally, we must include a term that describes the ``reduced'' AC Stark
shift $H_{pol}$ with strengths $\alpha_{\parallel}$ and $ \alpha_{\perp}$,
rank-2 tensor operators ${\cal O}_{\parallel}$ and ${\cal O}_{\perp}$,
and laser intensity $I$ \cite{Kotochigova2010}.  This Stark shift
is ``reduced'' in the sense that it is the Stark shift of the molecule
when the other terms in our Hamiltonian are ignored and the
``reduced'' polarizabilities $\alpha_{\parallel}$ and $\alpha_{\perp}$
only depend on the laser frequency. 
In fact, these two polarizabilities can be expressed in terms of a
sum over ro-vibrational states of all excited $^1\Sigma^+$ and $^1\Pi$ electronic potentials, respectively.
The operators ${\cal O}_{\parallel}$ and ${\cal O}_{\perp}$ capture all
dependence on light polarization and rotational angular momentum $\vec N$.
For a 1063 nm laser  $\alpha_{\parallel}/h= 10.0\times
10^{-5}$ MHz/(W/cm$^2$) and $\alpha_{\perp}/h=3.3\times 10^{-5}$
MHz/(W/cm$^2$)  measured in Ref.~\cite{Neyenhuis2012}.

We find eigen-energies of this Hamiltonian by diagonalization, including
rotational levels $N\le 20$, and analyze  its eigenfunctions to connect
to states that have been observed experimentally. Eigenstates can
be identified by the channel state with the largest contribution,
although for field strengths and laser intensities accessible in
ultracold molecular experiments we expect that the eigenstates can
be severely mixed.  
The polarizability of eigenstate $j$ with energy ${\cal E}_j(I,E)$ is
defined as the derivative $\alpha_j = -d{\cal E}_j/dI$, while the dipole
moment of state $j$ is ${\vec d}_j=-d{\cal E}_j/d{\vec E}$.  In this paper these two
quantities are studied as a function of angles $\theta$ and $\varphi$,
defined in Fig.~\ref{scheme}a, magnetic field strength $B$, electric
field strength $E$, and intensity $I$ of the trapping laser field.

\section{Real part of polarizability}\label{sec:reA}

In this section we present results for the dynamic polarizability of the
$N=0$ and $N=1$ rotational levels of the $v = 0$ vibrational level of the
ground X$^1\Sigma^+$ state of $^{40}$K$^{87}$Rb. The molecules are placed
in an optical dipole trap created from a focussed laser with a wavelength of 1063 nm.
An  external magnetic field of $B=545.9$ G and a static electric field are also present.
The values of the fields and laser intensities are based on recent measurements with
ultra-cold KRb molecules \cite{Neyenhuis2012}.

\begin{figure}
\begin{center}
\includegraphics[scale=0.78,trim=10 10 0 20,clip]{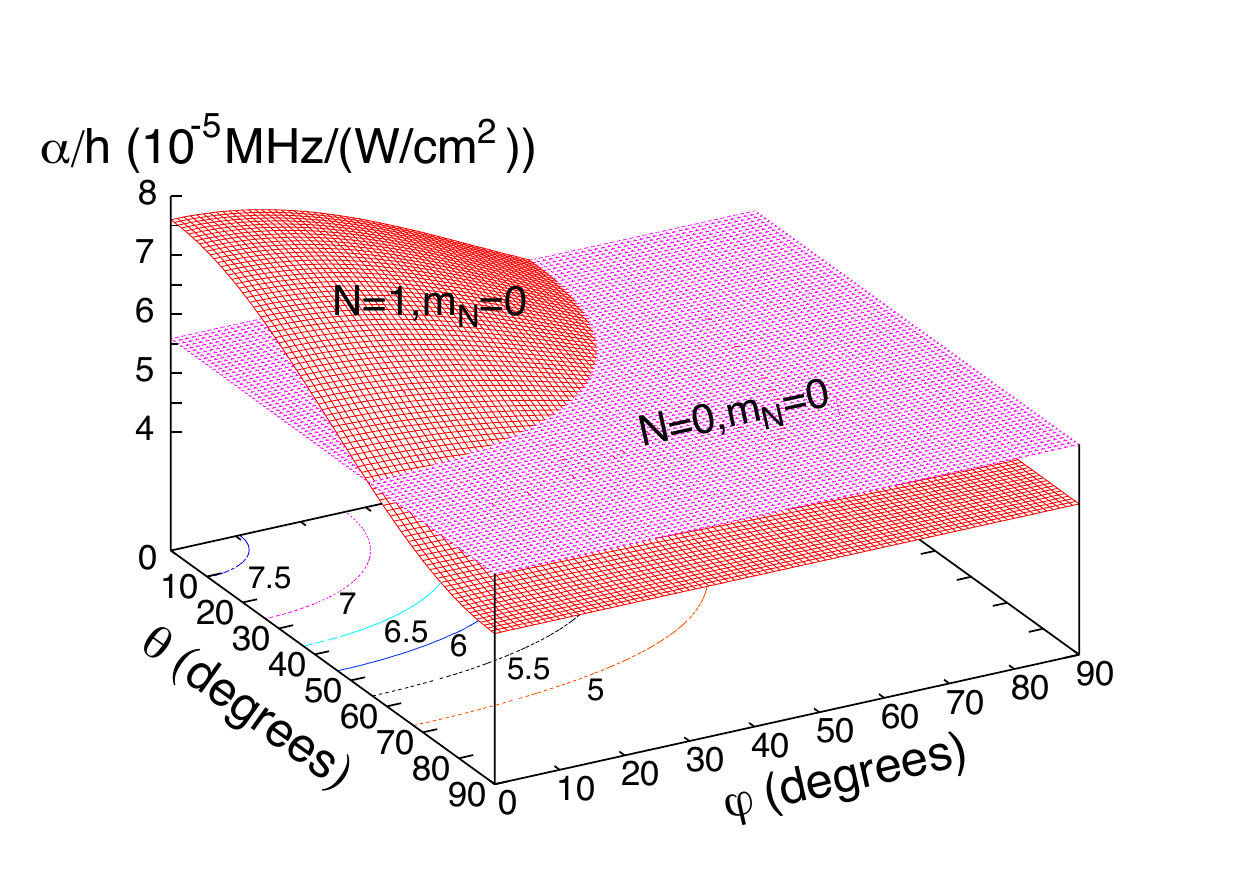}
\end{center}
\caption{(Color online) Polarizability of two $m_N=0$ rotational-hyperfine
states of the $v=0$ vibrational level of the X$^1\Sigma^+$ potential of
KRb as a function of angles $\theta$ and $\varphi$.  Six contours
of constant polarizability of the $|N=1,m_N=0\rangle$ state are plotted
as well. The polarizability of the $|N=0,m_N=0\rangle$ state is independent of
the two angles. The contour marked by 5.5 approximately corresponds to ``magic'' conditions
for the two $m_N=0$ rotational-hyperfine states.  
Parameters and remaining orientation are as for Fig.~\ref{polar_all}. 
 }

\label{polar00}
\end{figure}

Figure~\ref{polar_all} shows the polarizability as a function of
$\theta$ and $\varphi$ based on the  Hamiltonian in Eq.~\ref{eq:ham} 
and the geometry defined in Fig.~\ref{scheme}a for four
rotational-hyperfine states. The laser intensity $I=2.35$ kW/cm$^2$ and  the electric field
strength $E = 1$ kV/cm. 
We focus on the four states that have a predominant $|N, m_N, m_{\rm a}, m_{\rm b}\rangle= |0,0,-4,1/2\rangle
$, $|1,0,-4,1/2\rangle$, and $|1,\pm1,-4,1/2\rangle$ character as they are of experimental interest.
We observe that the polarizabilities of N=1 
states change noticeably when going from small to
and large values of the angles, while that for $|0,0,-4,1/2\rangle $ does not change
for any angle. Figure~\ref{polar_all} shows that the polarizabilities
of hyperfine levels coincide for many values of $\theta$
and $\phi$ angles. Crossings of polarizabilities correspond to the so-called 
``magic'' angles $\theta$ and $\varphi$, where the differential Stark
shift for two or more states is zero.  In fact, Fig.~\ref{polar00} shows that
the magic angles between the states $|0,0,-4,1/2\rangle $ and $|1,0,-4,1/2\rangle $ 
form a nearly circular, elliptical curve that starts at $\theta=57$ degrees and $\varphi=0$ degrees.

\begin{figure}
\begin{center}
\includegraphics[scale=0.33,trim=0 25 0 80,clip]{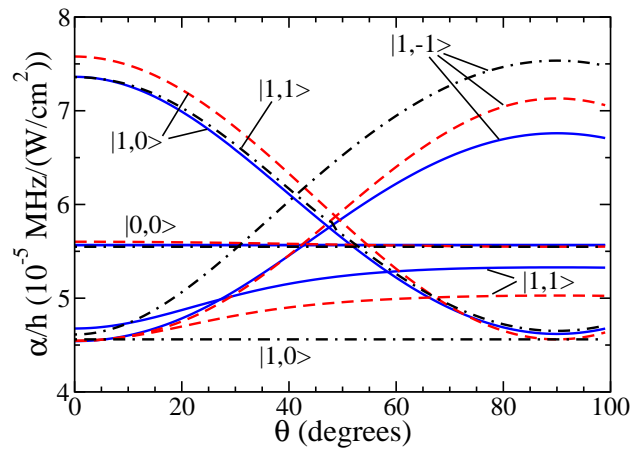}
\end{center}
\caption{(Color online) Polarizability of four rotational-hyperfine states of the $v=0$ vibrational
level of the X$^1\Sigma^+$ potential of KRb as a function of angle $\theta$ between the polarization 
of the dipole-trap laser for three configurations of the electric field.
The solid blue lines are for zero electric field. They were previously published in Ref.~\cite{Neyenhuis2012}. 
The red dashed lines are for  an electric field with a strength of 1 kV/cm oriented parallel
to the magnetic field ($\varphi=0^{\rm o}$), while the dash-dotted black lines are for  $E = 1$ kV/cm oriented in the 
$y$-$z$ plane but perpendicular to the magnetic field ($\varphi=90^{\rm o}$).
Remaining parameters and nuclear hyperfine states are as for Fig.~\ref{polar_all} and note that the $x$ axis extends to $100^{\rm o}$. 
}
\label{polar_magic}
\end{figure}

\begin{figure}[b]
\begin{center}
\includegraphics[scale=0.33,trim=0 25 0 80,clip]{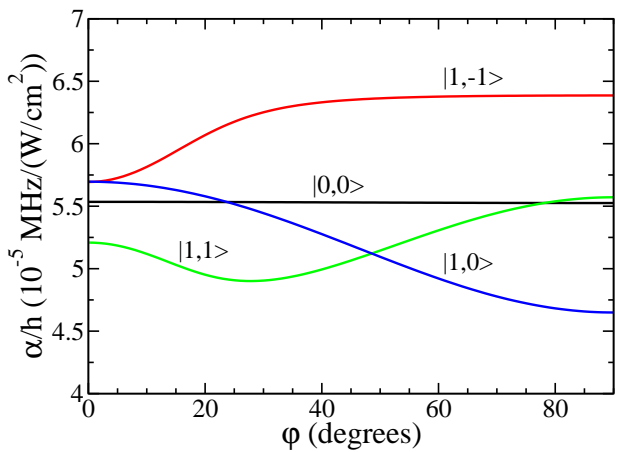}
\end{center}
\caption{(Color online) Polarizability of four rotational-hyperfine states of the lowest vibrational
level of the X$^1\Sigma^+$ potential of KRb as a function of angle $\varphi$ between the electric field and
the magnetic field and $\theta=51$ degrees. For this value of $\theta$ the ``magic'' angle $\varphi$ is 23 degrees.  
All other parameters are the same as for Fig.~\ref{polar_all}.
}
\label{polar_phi}
\end{figure}

Figures \ref{polar_magic} and \ref{polar_phi} show cuts through the surfaces depicted in Figs.~\ref{polar_all} and
\ref{polar00} in order to facilitate a quantitative comparison. Figure \ref{polar_magic} shows the polarizability
for four rotational-hyperfine states as a function of $\theta$ without applied electric field as well as for $E=1$ kV/cm oriented 
either parallel ($\varphi=0^{\rm o}$) and perpendicular ($\varphi=90^{\rm o}$) to the magnetic field. The curves for 
zero electric field and that for $E=1$ kV/cm with $\varphi=0^{\rm o}$ are similar in shape and predict ``magic'' conditions
between $\theta=50^{\rm o}$ and $60^{\rm o}$.  The polarizability for
the  $|1,0,-4,1/2\rangle$ state decreases by as much as a factor of two for increasing $\theta$.
On the other hand, for $E=1$ kV/cm and $\varphi=90^{\rm o}$ the polarizabilities of $|0,0,-4,1/2\rangle$ and $|1,0,-4,1/2\rangle$ do not cross. In fact,  the polarizability of the $|1,0,-4,1/2\rangle$ state is nearly independent of angle $\theta$, while that for the $|1,1,-4,1/2\rangle$ state now decreases by a factor of two for increasing $\theta$.

Figure~\ref{polar_phi} shows the polarizability  for the same four states as a function of $\varphi$ for one value of $\theta$ and $E=1$ kV/cm. 
For this value of $\theta$ the magic condition between states $|0,0,-4,1/2\rangle$ and $|1,0,-4,1/2\rangle$ occurs at the
relatively-small angle $\varphi=23^{\rm o}$. On the other hand, the polarizability of the $|1,-1,-4,1/2\rangle$ state does not coincide with that of the
$|0,0,-4,1/2\rangle$ state at any angle $\varphi$.

Our analyses shows that for a trapping laser light at 1063 nm and external
electric field strengths of 1 kV/cm only a few low-lying rotational
states are mixed.  The near infrared laser frequency is detuned away from resonances
with rovibrational levels of the electronically excited potentials. As
a result, corrections to the polarizability from the level shifts due
the static electric and magnetic field are significantly suppressed.

\section{Induced dipole moment}\label{sec:dip}

\begin{figure}
\begin{center}
\includegraphics[scale=0.33,trim=0 15 0 80,clip]{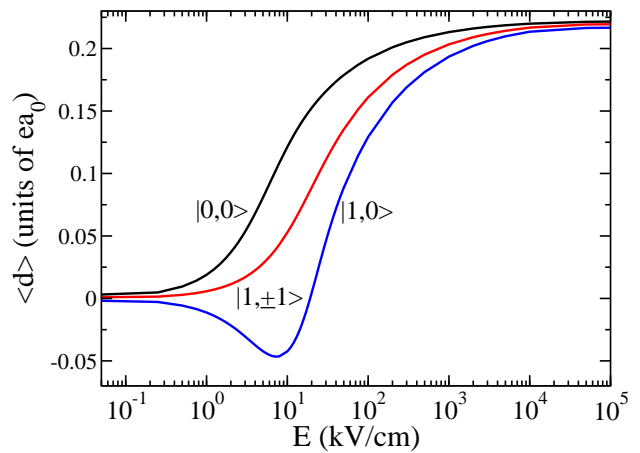}
\end{center}
\caption{(Color online) The induced dipole moment of  the lowest rotational levels of the $v=0$ vibrational 
level of the X$^1\Sigma^+$ potential of $^{40}$K$^{87}$Rb as a function of external electric field strength $E$. 
The angles $\theta$ and $\varphi$ are  51 degrees and 23 degrees, respectively. The laser intensity, remaining orientation, and  magnetic field strength are the same as for Fig.~\ref{polar_all}. }
\label{dipmom}
\end{figure}

Figure~\ref{dipmom} shows the induced dipole moment of four
rotational-hyperfine levels along the electric field direction  as a
function of the electric field strength $E$. We assume an angle $\theta=
51^{\rm o}$ between the magnetic field and the laser polarization and
angle $\varphi= 23^{\rm o}$ between the electric and magnetic fields,
to ensure that the $|N,m_N\rangle=|0,0\rangle$ and $|1,0\rangle$ states
have the same polarizability at $E=1$ kV/cm. The nuclear spin state is
$m_{\rm a}=-4$ and $m_{\rm b}=1/2$.  For $E < 7$ kV/cm the dipole moment
of the  $|1,0\rangle$ state decreases with $E$ and is negative.  On the
other hand, dipole moment of $|0,0\rangle$ and $|1,\pm1\rangle$  states
always increase with $E$ and are positive.  For  $E\gg 10$ kV/cm the
induced dipole moments of all four rotational-hyperfine levels converge to
$d_0=0.223\, ea_0$.  From results not shown here, we find that the induced
dipole moment is aligned along the direction of the electric field, ${\vec
d}_j\propto \vec E$, and that the magic conditions are nearly independent
of electric field strength. Both observations follow from the fact that
$m_{\rm a}=-4$ and $m_{\rm b}=1/2$ are approximately good quantum numbers
and that, for fields shown in Fig.~\ref{dipmom},  $H_E$ is much larger
than $H_Z$ and $H_Q$. The coupling between rotational and nuclear spin
states is weak.  Then, for small electric field strengths, the four
${\vec d}_j$ follow the second-order perturbation theory expressions
\begin{equation}
\begin{array}{l}
\displaystyle
  \vec d_{N,m_N=0,0} = 2 \frac{d_0^2|\langle 0,0|C_{10}|1,0\rangle|^2}{2B_v} \vec E; \\
\displaystyle
  \vec d_{1,0} = 2\left[- \frac{d_0^2|\langle 0,0|C_{10}|1,0\rangle|^2}{2B_v} +  
                       \frac{d_0^2|\langle 2,0|C_{10}|1,0\rangle|^2}{4B_v}\right] \vec E; \\
\displaystyle
 \vec d_{1,\pm1} = 2 \frac{d_0^2|\langle 2,\pm1|C_{10}|1,\pm1\rangle|^2}{2B_v} \vec E\, .
\end{array}
\label{perttheory}
\end{equation}
The induced dipole moment of the $|N,m_N\rangle=|0,0\rangle$ and
$|N,m_N\rangle=|1,\pm 1\rangle$ states have a single contribution
from transitions to states with $N'=N+1$ with larger rotational
energies. This leads to a positive dipole moment. For the dipole moment
of the $|N,m_N\rangle=|1,0\rangle$  state contributions from state with
both smaller and larger rotational energies appear. In this case their
combined effect leads to a negative dipole moment.

\section{Imaginary part of the polarizability}\label{sec:ImA}
 
The Hamiltonian described in Section \ref{sec:ham} does not describe
losses due to spontaneous emission of ro-vibrational levels of
electronically excited states. These losses appear as an imaginary
contribution to the polarizability. We can understand this by realizing that
at a specific laser intensity, magnetic and electric field
the complex-valued polarizability of state $i$ can also be defined as
\begin{eqnarray}
   \lefteqn{ \alpha(h\nu,\vec{\epsilon}) = } 
      \label{eqpolar}
\\
  &&\frac{1}{\epsilon_0c}
   \sum_{f} \frac{(E_f - ih\gamma_f/2 - E_i)}{(E_f - ih\gamma_f/2 - E_i)^2 - (h\nu)^2}
     \times |\langle f|{\vec d_{tr}} \cdot \vec{\epsilon}|i\rangle|^2\,, \nonumber
\end{eqnarray}
where $c$ is the speed of light, $\epsilon_0$ is the electric constant,
and kets $|i\rangle $ and $|f\rangle$ denote initial and intermediate
rotational-and-hyperfine-resolved vibrational wavefunctions of the
$|X^1\Sigma^+\rangle$ potential and excited electronic states. Their
energies are $E_i$ and $E_f$, respectively.  The matrix elements $\langle
f|\vec d_{tr}|i \rangle$ are vibrational-averaged electronic transition
dipole moments and $\vec \epsilon$ is the polarization of the laser.  The sum
over $f$ excludes the initial state but includes transitions to the
rovibrational levels within the X$^1\Sigma^+$ potential as well as to
the rovibrational levels of excited potentials.  Contributions from
scattering states or the continuum of any state must also be included.
For alkali-metal dimers this sum, however, can be limited to transitions
to electronic excited potentials  that dissociate to
either a singly-excited K or Rb atom as only those have significant
electronic dipole moments to the  X$^1\Sigma^+$ state.  Moreover, as we focus on the KRb polarizability
for an infrared laser with a 1063 nm wavelength, their contribution is
further reduced by the energy denominator in Eq.~\ref{eqpolar}.
Finally, the natural line widths $\gamma_f$
of excited ro-vibrational levels describe the spontaneous emission that
lead to loss of molecules by emission of a spontaneous photon.

As currently only a single measurement of the imaginary part of the polarizability
\cite{Chotia2012} is available to characterize the imaginary part of
the two ``reduced'' polarizabilities of $H_{pol}$ in Eq.~\ref{eq:ham},
we can only compare this $\vec E=\vec 0$ measurement to {\it ab-initio} theoretical
estimates based on Eq.~\ref{eqpolar}.  To calculate the theoretical dynamic polarizability
we use the most accurate ground state potentials available from
Ref.~\cite{Tiemann}.  Excited potentials are  constructed from RKR
data \cite{Kasahara,Amiot1} and as well as from our {\it ab~initio}
calculations \cite{Kotochigova2009} using long range dispersion
coefficients from Ref.~\cite{Bussery}.  We employ transition dipole
moments from our previous electronic structure calculations of KRb
\cite{Kotochigova2003, Kotochigova2004}.

\begin{figure}
\begin{center}
\includegraphics[scale=0.33,trim=0 25 0 40,clip]{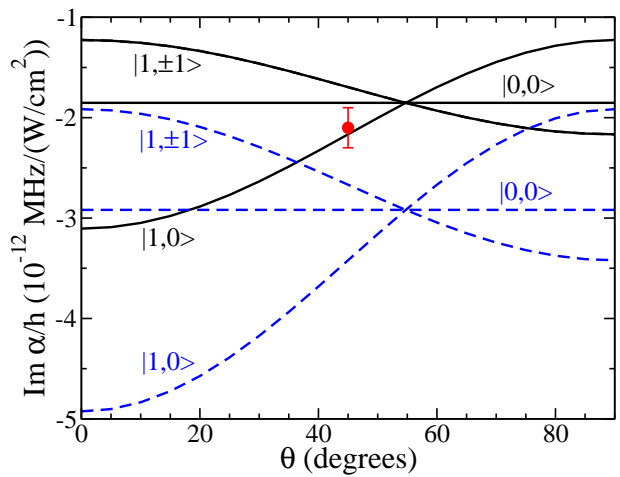}
\end{center}
\caption{(Color online) Imaginary part of polarizability of $N=0$ and 1 rotational levels of the $v=0$
vibrational level of the X$^1\Sigma^+$ potential of KRb as a function of
angle $\theta$ between the polarization of the dipole-trap laser at 1063
nm and bias magnetic field with strength $B=545.9$ G. The static electric
field strength is zero. The laser has an intensity of  $I=2.35$ W/cm$^2$.
Results for two different ways of determining the imaginary polarizability
are shown. The solid lines correspond to ${\rm Im}\, \alpha$, when the linewidth
of excited ro-vibrational levels equals the atomic linewidth of K, the
dashed lines are obtained using excited state linewidths obtained with
an ``optical potential'' approach. The only experimental measurement of
the imaginary polarizability for the $|N, m_N\rangle=|0,0\rangle$ state and
$\theta$ = 45 degrees  \cite{Chotia2012} is shown by the red marker with
error bar.}

\label{impol}
\end{figure}

We evaluate the linewidths of excited rovibrational states using
two different methods.  In the first method, the imaginary part of the
polarizability is calculated assuming that the linewidth of rovibrational
levels of the X$^1\Sigma^+$ potential is zero and that rovibrational
levels of the lowest excited electronic potentials that dissociate to
either a singly-excited K or Rb atom, have a natural linewidth equal to
the atomic linewidth of potassium. The small, less than 1\% difference
in linewidth of K and Rb does not modify our results significantly.

For our second method the molecular line width of vibrational
levels of potentials that dissociate to either a singly-excited K or
Rb atom is calculated assuming an ``optical potential'' $i\Gamma(r)/2$
\cite{Zygelman}, where $\Gamma(r)$ is proportional to $\omega(r)^3
d(r)^2$, the frequency $\omega(r)$ is the transition frequency between the
two potentials at each $r$, and $d(r)$ is $r$-dependent transition dipole
moment. This is essentially a stationary phase approximation.  The number
of electronic potentials included is the same as for the first method.

Figure~\ref{impol} shows the calculated imaginary part of the
polarizability as a function of angle $\theta$ for the two means of
including the effect of spontaneous emission.  The value of the imaginary
part is always negative and is seven orders of magnetic smaller than
the real part.  Eventhough this imaginary part is small, it will affect
precision measurements with ultracold KRb molecules.  The figure also shows that 
the absolute value of Im $\alpha$ is larger for the second method of modeling spontaneous emission.  
This is because the molecular  transition dipole moments
from the excited state to the ground state  at the equilibrium separation
$R_e$ is larger than the atomic dipole moment.  The measured
${\rm Im}\, \alpha$ = $-2.1(2)\times10^{-12}$ MHz/(W/cm$^2$) for the $|N,
m_N\rangle=|0,0\rangle$ state and $\theta$ = 45 degrees \cite{Chotia2012}
is in  better agreement with the model that uses the atomic linewidth.

\section{Summary}
We performed a  theoretical study of  the internal
rovibronic and hyperfine quantum states of the KRb molecules when
simultaneously static magnetic and electric fields as well as trapping
lasers are applied.  The combined action of these field can be used
for an efficient quantum control of ultracold polar molecules in
optical potentials. We extended the ideas of mixing rotational levels in
Refs.~\cite{Kotochigova2010,Neyenhuis2012} to include all three fields as
in typical ultracold experiments. In particular, we searched for ``magic''
angles between external DC electric, magnetic, and AC trapping fields,
where the  AC Stark shift of pairs of rotational states are the same. Moreover,
we evaluated the induced dipole moment of the internal
rovibronic and hyperfine quantum as a function of external electric field. 
With this precise value of the dipole
moment one can investigate of how interactions between
molecules in the different optical lattice sites depend on the relative
orientation of the applied fields.  Our theoretical research efforts
are closely linked to ongoing experiments with ultracold KRb molecules.

\section*{Acknowledgments}

We acknowledge funding from  ARO MURI Grant W911NF-12-1-0476 
and NSF Grants NSF PHY-1005453 and NSF PHY11-25915.

\end{document}